\documentclass[reprint,amsmath,amssymb,prd,superscriptaddress,floatfix]{revtex4-2}
\usepackage[T1]{fontenc}
\usepackage{slashed}
\usepackage{subcaption}
\usepackage[utf8]{inputenc} 
\usepackage{graphicx}
\newcommand{\be}{\begin{equation}}
\newcommand{\ee}{\end{equation}}

\makeatletter
\def\@fnsymbol#1{\ensuremath{\ifcase#1\or *\or  \ddagger\or \mathsection\or \|\or
    \mathparagraph\or \|\or **\or \P\dagger
   \or \ddagger\ddagger \else\@ctrerr\fi}}
\makeatother

\begin{document} 

\title{Fundamental Landau Levels in a Strongly Coupled Plasma.}

\author{Uriel Elinos}
 \email{elinos@ciencias.unam.mx}
 \affiliation{Departamento de F\'{\i}sica, Facultad de Ciencias, Universidad Nacional Aut\'onoma de M\'exico,\\Apartado Postal 70-542, CP 04510, Ciudad de M\'exico, M\'exico.}
  \affiliation{Departamento de F\'{\i}sica, Universidad Aut\'onoma Metropolitana—Iztapalapa, Avenida San Rafael Atlixco 186, A.P. 55534, C.P. 09340, Ciudad de M\'exico, M\'exico.}
\author{Leonardo Pati\~no}
 \email{leopj@ciencias.unam.mx}
 \affiliation{Departamento de F\'{\i}sica, Facultad de Ciencias, Universidad Nacional Aut\'onoma de M\'exico,\\Apartado Postal 70-542, CP 04510, Ciudad de M\'exico, M\'exico.}

\begin{abstract}
We use the gauge/gravity correspondence to show that in its context there appear Landau levels when studying the excitations of the fundamental degrees of freedom of a strongly coupled plasma subject to a magnetic field of arbitrary intensity. We work in the phase where mesons are melted by embedding a D7-brane in the 10D uplift that we construct for the 5D background associated to a magnetic brane. By studying the dependence of the lowest quasinormal frequency on both, the intensity of the magnetic field and the Landau level, we corroborate that the energy behaves as expected for dissociating Landau levels. The reconstruction of the impact of the magnetic field on the stability of the quasinormal modes allows us to identify inverse magnetic catalysis for intensities of the field bellow a certain value, and magnetic catalysis for those above it.
\end{abstract}


\maketitle

\section{Introduction and main results}
\label{sec:intro}

The state of matter, known as Quark-Gluon plasma (QGP), produced in high energy collisions like those carried on in the Relativistic Heavy Ion Collider (RHIC) or in the Large Hadron Collider (LHC), currently represents one of our most important windows to look at the properties of strongly coupled Quantum Chromodynamics (QCD) and to explore many other characteristics of fundamental physics.

The progress in understanding the properties of the QGP has been achieved by systematically analyzing the patterns of particles emitted by the plasma instantly after its creation. Theoretical tools that use our current understanding of strong interactions are necessary to determine which phenomena is consistent with the data compiled from the detectors, making it crucial for our theory to be as accurate as possible, both, in its foundations as well as having taken into account any relevant ingredient.

In this last sense, it seems to be clear that the presence and transcendence of a strong magnetic field in the QGP is undeniable \cite{Skokov:2009qp,Wilde:2012wc,Basar:2012bp,Andersen:2014xxa,Ayala:2018wux}. The system, being strongly coupled, escapes the strict applicability of perturbative strategies in quantum field theory, so it has to be treated using non-perturbative methods which for the problems at hand are not fully developed. Another alternative is the use of lattice computations, but their euclidean nature limits the reach of the analysis. A perspective that has been extensively applied is the one provided by the gauge/gravity correspondence, which even if it does not describe Quantum Chromodynamics, it deals with a strongly coupled theory with some properties that approach those of QCD in certain limits, so we can hope that some observables are robust enough to be extrapolated from one theory to the other, at least qualitatively. In particular, the authors in \cite{DHoker:2009mmn} constructed a five dimensional background to model a dual strongly coupled plasma immersed in a magnetic field of arbitrary intensity.

Even if some phenomena can be robust enough in the sense that we mentioned before, a lot of effort has also been invested to bring the gauge theories that can be described by the correspondence to be as similar as possible to QCD. One of these improvements is the inclusion of matter fields in the fundamental representation, as they exist in real world Physics, done by embedding probe branes, referred to also as flavor branes, in geometries with dual theories that contain only fields in the adjoint representation \cite{Karch:2002sh,Kruczenski:2003be,Babington:2003vm}. In the present work we begin by performing this extension for the family of backgrounds developed in \cite{DHoker:2009mmn}, which requires as a first step the construction of the ten dimensional uplift of the latter, to then proceed with the embedding of D7-branes in it. As we will see, the dual gauge theory in which we end up working is four dimensional, contains fields in the fundamental representation, is at finite temperature, and is subjected to a constant magnetic field $F_{GT}=b\,dx\wedge dy$ of arbitrary intensity.

Our construction indicates that the peculiarities of the 10D uplift demand the D7-branes to develop an embedding profile that is not trivial in the directions of the dual gauge theory, except when it lies at the equator of the compact space and the profile is unaltered with respect to its supersymmetric equivalent in the $\{T\rightarrow 0, b\rightarrow 0\}$ limit. A main result that we present is that one of the channels of the excitations of the strongly coupled plasma dual to said equatorial embedding are governed in the gauge theory directions by an equation identical in form to the one leading to Landau quantization in non-relativistic Quantum Mechanics. The last statement is always true when using Landau gauge for the U(1) vector potential $A=b\,x\,dy$ in the construction of the uplift, provided that the origin of the coordinate system is adjusted so that the guiding center parameter is equal to zero. In the symmetric gauge given by $A=b(\,x\,dy-y\,dx)/2$, we find that, as a system, the two equations that dictate the profile are equivalent to the one that describes Landau levels in ordinary Quantum Mechanics for states of vanishing canonical angular momentum. It is important to notice that the latter restriction does not imply the vanishing of the kinetic angular momentum, for which we provide an spectrum.

The derivation of everything that has been described up to this point, can be done based only on the general form of the 10D uplift that we obtain for the family of backgrounds constructed in \cite{DHoker:2009mmn}. The specific profile of the perturbation of the D7-brane embedding in the radial direction of the asymptotically AdS space is determined by the explicit dependence of the background metric on this coordinate. We refer to appendix A in \cite{Arean:2016het} and provide just a brief explanation about how to numerically construct this metric, which we then use to build the aforementioned radial perturbation profile. A first objective of doing this is to prove that the D7-brane remains perturbatively close to the equator, validating the analysis of the excitations.

The identification of the discrete set of on shell perturbations characterized by a normalizable radial profile permits us to study the quasinormal modes of the dual scalar excitations. We present the dependence of the lowest quasinormal frequency on both, the intensity $b$ of the magnetic field and the Landau level number, from which we extract the physical effect that these have on the energy $E$ and width $\Gamma$ of the dual state. We show that as a function of $b$, the energy of our sates is very close to the one of fully stable Landau levels, and present this as evidence that the evolution of the modes we found describes the process of dissociation of the latter.

When $\Gamma/E$ is displayed as a function of $b/T^2$ we observe that the states become more stable as this ratio begins to grow from zero, continuing up to a critical value $b_c$ that depends on the level, and after which the stability decreases with the increment of $b/T^2$. This coincides with the inverse magnetic catalysis and magnetic catalysis for different ranges of the magnetic field reported in \cite{Ayala:2014yla}. 

It is worth mentioning that, to the top of our knowledge, prior studies of Landau levels using a background that includes a magnetic field in the context of the gauge/gravity correspondence had been done either introducing probe fermions in such background \cite{Albash:2009wz,Albash:2010yr,Blake:2012tp}, or perturbing them directly to find their excitation modes \cite{Ammon:2016fru,Ammon:2017ded,Ammon:2020rvg}, and hence not including fields in the fundamental representation of the dual gauge theory. In contrast, fundamental degrees of freedom were introduced in \cite{Kristjansen:2012ny,Kristjansen:2013hma,Hutchinson:2014lda} through the inclusion of D5-branes and D7-branes, but in those instances the backgrounds where the latter were embedded did not bear the magnetic field, which was incorporated as a worldvolume gauge field, making the particular constructions only sensitive to $F=d\,A$, and therefore unsuitable for the detail treatment of Landau quantization which involves $A$ directly.

\section{Magnetic branes and their ten dimensional uplift}
\label{sec:Uplift}
The five dimensional metric of the background constructed in \cite{DHoker:2009mmn} to subject the dual strongly coupled plasma to a constant magnetic field $F_{GT}=b\,dx\wedge dy$ can be conveniently cast in the form
\begin{equation}
ds_{_5}^2 = -U(r)dt^2+\frac{dr^2 }{U(r)} + V(r)\left( dx^2 + dy^2\right) + W(r) dz^2.\label{ds5}
\end{equation}

In writing \eqref{ds5}, all the expected symmetries of a geometry generated by a D3-brane had been taken into account, except those broken by a constant magnetic field pointing in the $dx\wedge dy$ direction. For this metric and gauge field to form the solution to 5D minimal gauged supergravity that we would like them to be, they have to solve the equations that result from the variation of
\be
\begin{split}
&S=\frac{1}{16\pi G_5}\Big[-\int d^5x\sqrt{-g}\Big(R+F^{MN}F_{MN}-\frac{12}{L^2}\Big)\\
&+\frac{8}{3\sqrt{3}}\int A\wedge F\wedge F\Big],
\end{split}
\label{S5}
\ee
which is the bosonic part of the action that corresponds to the aforementioned theory with $F=dA$, and where none of the possible boundary terms have been included because they will not be relevant for our present calculation. In all that follows, including the ten dimensional treatment, we will fix the AdS radius $L$ to 1.

Once the elements of \eqref{ds5} have been inserted in the equations that arise from \eqref{S5}, it is easy to see that those coming from the variation with respect to $A$ are solved by $F=Bdx\wedge dy$, while the ones associated to the variation with respect to the metric reduce to
\be
\begin{split}
&W(r) \left[-8 B^2+6 V(r) \Big( V'(r)U'(r)+ V(r)U''(r) \Big)\right]\\
&+ 3 V(r)^2 \Big( U'(r) W'(r)-16 W(r)\Big)=0,\\
&2 W(r)^2 \,\Big[4 B^2  +V(r)\, \Big(U'(r)\, V'(r)+U(r) \, V''(r)\Big)\Big]\\
&+U(r) V(r)^2 W'(r)^2-\, V(r)\, W(r)\,\, \Big[U(r) V'(r) W'(r)\\
&+2 V(r)\Big( U'(r) W'(r)+U(r) W''(r)\Big)\Big]=0,\\
&4 V(r) W(r)^2 V''(r)-2 W(r)^2 V'(r)^2\\
&-V(r)^2\Big( W'(r)^2-2 W(r) W''(r)\Big) = 0,  \\
& W(r) \Big(4 B^2+2 V(r) U'(r) V'(r)+U(r) V'(r)^2 -24 V(r)^2\Big)\\
&+V(r) W'(r)\Big(V(r) U'(r)+2 U(r) V'(r)\Big) = 0.
\end{split}\label{eqback}
\ee

In the last paragraph we used $B$ instead of $b$ for reasons that will become clear in section \ref{solution}, where we will see a way to build a family of black D3-branes solutions with a horizon located at $r_h$, that are asymptotically AdS5, have a common temperature $T=U'(r_h)/4\pi$, and are parametrized by the intensity $b$ of the magnetic field which can take any value between 0 and infinity.

What we are currently concerned with is the inclusion of fundamental degrees of freedom in the plasma dual to the configurations that we have just described, and hence require their uplift to ten dimensions. Once in possession of such 10-D backgrounds, we will embed D7-branes so that fields in the fundamental representation are part of the dual theory.

It was shown in \cite{Cvetic:1999xp} that the general equations that follow from \eqref{S5} for the five dimensional metric and $U(1)$ field intensity, can also be obtained as the reduction of those of D=10 type IIB supergravity by writing the ten dimensional line element in terms of $ds_5^2$ and the former gauge potential $A$ as
\be
\begin{split}
ds_{10}^2&=ds_5^2+ \bigg\{  d\theta ^{2}+\sin^{2}\theta \left(d\phi_{1}+\frac{2}{\sqrt{3}}A_\mu dx^\mu \right)^{2}   \\
&  + \cos^{2}\theta \bigg[ d\vartheta^{2} +\sin^{2}\vartheta \left(d\phi_{2}+\frac{2}{\sqrt{3}}A_\mu dx^\mu \right)^{2}\\
& +\cos^{2}\vartheta \left(d\phi_{3}+\frac{2}{\sqrt{3}}A_\mu dx^\mu \right)^{2}\bigg] \bigg\},
\end{split}\label{ds10}
\ee
and the 5-form as
\be
F_{(5)}=G_{(5)}+ *G_{(5)},\label{F5}
\ee
with
\be
\begin{split}
G_{(5)}&=-4\epsilon_{(5)}+\frac{2}{\sqrt{3}} \left[ \sin\theta \cos\theta d\theta \wedge \left(d\phi_{1}+\frac{2}{\sqrt{3}}A_\mu dx^\mu \right)\right. \\ 
&-\cos\theta\sin\theta \sin^{2}\vartheta d\theta \wedge \left(d\phi_{2}+\frac{2}{\sqrt{3}}A_\mu dx^\mu \right)\\
&+ \sin\vartheta\cos\vartheta\cos^{2}\theta d\vartheta \wedge \left(d\phi_{2}+\frac{2}{\sqrt{3}}A_\mu dx^\mu \right) \\
& - \cos\theta \sin\theta \cos^{2}\vartheta d\theta \wedge \left(d\phi_{3}+\frac{2}{\sqrt{3}}A_\mu dx^\mu \right)\\
&\left. - \cos\vartheta\sin\vartheta\cos^{2}\theta d\vartheta \wedge \left(d\phi_{3}+\frac{2}{\sqrt{3}}A_\mu dx^\mu \right) \right]\wedge\bar{*}F,
\end{split}
\ee
where $\epsilon_{(5)}$ and $\bar{*}$ are the five dimensional volume element and Hodge star operator respectively.

It is a simple but relevant consistency check to verify that once \eqref{ds5} and $F=Bdx\wedge dy$ have been explicitly placed in \eqref{ds10} and \eqref{F5}, the substitution of the resulting
{\small
\be
\begin{split}
&ds_{10}^2=-U(r)dt^{2} + \frac{1}{U(r)}dr^{2}+V(r) \left( dx^{2} + dy^{2} \right)+ W(r)dz^{2} \\
&+ \left[ d\theta ^{2}+\sin^{2}\theta d\widetilde{\phi}_1^2 + \cos^{2}\theta\left( d\vartheta^{2} +\sin^{2}\vartheta d\widetilde{\phi}_2^2+\cos^2\vartheta d\widetilde{\phi}_3^2\right)  \right],    
\end{split}\label{ds10ds5}
\ee}
and
{\small
\be
\begin{split}
&F_{(5)}=-4\sqrt{V^{2}W}\, dt \wedge dr \wedge dx \wedge dy \wedge dz \\
&+\frac{2B\sqrt{V^{2}W}}{V^{2}\sqrt{3}}  dt \wedge dr \wedge dz \wedge \left[ \sin\theta \cos\theta\, d\theta \wedge d\widetilde{\phi}_{1} \right.  \\ 
&+(-\cos\theta\sin\theta \sin^{2}\vartheta \, d\theta + \sin\vartheta\cos\vartheta\cos^{2}\theta \, d\vartheta) \wedge d\widetilde{\phi}_{2}  \\
&\left. +(-\cos\theta \sin\theta \cos^{2}\vartheta \, d\theta    - \cos\vartheta\sin\vartheta\cos^{2}\theta \, d\vartheta )\wedge d\widetilde{\phi}_{3}  \right] \\
&-4\cos^{3}\theta \sin\theta \sin\vartheta \cos\vartheta d\theta \wedge d\vartheta \wedge d\widetilde{\phi}_{1} \wedge d\widetilde{\phi}_{2} \wedge d\widetilde{\phi}_{3}\\ 
&- \frac{2B}{\sqrt{3}}dx \wedge dy\wedge \left[  \cos^{4}\theta \sin\vartheta \cos\vartheta \, d\vartheta \wedge d\widetilde{\phi}_{2} \wedge d\widetilde{\phi}_{3} \right. \\
&+(\cos\theta \sin\theta \cos^{2}\vartheta \, d\theta + \cos^{2}\theta \sin^{2}\theta \sin\vartheta \cos\vartheta \, d\vartheta) \wedge d\widetilde{\phi}_{1} \wedge d\widetilde{\phi}_{3}  \\
&\left. +(\cos\theta \sin\theta \sin^{2}\vartheta \, d\theta-\cos^{2}\theta \sin^{2}\theta \sin\vartheta \cos\vartheta \, d\vartheta   )\wedge d\widetilde{\phi}_{1} \wedge d\widetilde{\phi}_{2} \right],    
\end{split}\label{F5ds5}
\ee}into the equations of type IIB supergravity in ten dimensions indeed leads to \eqref{eqback} and nothing else, regardless of whether we use $d\widetilde{\phi}_i=d\phi_i+\frac{2}{\sqrt{3}}B\, x\, dy$ or  $d\widetilde{\phi}_i=d\phi_i+\frac{1}{\sqrt{3}}B(x\,dy-y\,dx)$, which correspond to two different gauge choices for $A$. The details of this calculation are not enlightening, so we do not include them, but once we carried it out, we are certain that any solution to \eqref{eqback} can be turned into a solution to 10-D type IIB supergravity through \eqref{ds10} and \eqref{F5}.

In view of the above we can assert that \eqref{ds10ds5} and \eqref{F5ds5} constitute the uplift of the 5 dimensional background constructed in \cite{DHoker:2009mmn} for the metric functions $U, V,$ and $W$ that can either be computed from those presented there, or in the manner we will develop in section \ref{solution}.

\section{Embedding of the D7-brane}

We are now in a position to incorporate fundamental degrees of freedom in the dual gauge theory, and for them to propagate in all of its directions, we will do it by embedding a probe D7-brane that extends along the non-compact directions $\{t,r,x,y,z\}$ of the 10-D background that we just constructed, and wraps a 3-cycle of its compact space $\Omega_5$, which line element is given by the second line of \eqref{ds10ds5}. The way in which the D7 lies in the ten dimensional space is then given by the manner in which the 3-cycle is accommodated in $\Omega_5$, and pursuing stability, we do it so that supersymmetry is recovered in the $B\rightarrow 0, T\rightarrow 0$ limit, where our configurations will reduce to the supersymmetric ones build in \cite{Karch:2002sh}. It will also be true that if we keep a finite temperature and set $B=0$, the construction done in \cite{Kruczenski:2003be,Babington:2003vm,Mateos:2007vn,Mateos:2007yp} would be recovered from ours.

Given the field content that we are considering, the D7-brane is only sensitive to the background metric, allowing us to place it based on geometric considerations alone. Specifically, the embedding is dictated by the DBI action
\be
S_{DBI}=-T_{D7}N_{f}\int d^{8}x\sqrt{-\text{det}(g_{D7})},\label{dbi}
\ee
where $g_{D7}$ is the induced metric on the D7-brane, and its tension is given by
\be
T_{D7}=\frac{1}{(2\pi l_{s})^{7}l_{s}g_{s}}=\frac{1}{16\pi^{6}}\lambda N_{c},\label{Tension}
\ee
while the constants $N_c$ and $N_f$ respectively indicate the number of D3-branes sourcing the background and D7-branes embedded in it. The probe approximation for the D7-brane is maintained by keeping $N_f\ll N_c$.

To find a suitable way to describe the positioning of the D7-brane, we begin by noticing that in the supersymmetric limit  $B\rightarrow 0, T\rightarrow 0$, the metric of $\Omega_5$ given by the second line in \eqref{ds10ds5} reduces to that of a 5-sphere of radius $L=1$, written as the product of, from right to left, a 3-sphere in Hopf (toroidal) coordinates, a circle in $\phi_1$, and a polar coordinate $\theta$ between them.

A supersymmetric embedding of a D7-brane in this limiting background is achieved \cite{Karch:2002sh} by letting the 3-cycle that it wraps in $\Omega_5$ coincide with the 3-sphere of the factorization above, keep it localized at a constant $\phi_1=\phi_1^0$, and placing it in the $r-\theta$ plane so that over its worldvolume the relationship $r\sin\theta=R$ is satisfied by $r$ and $\theta$ for some constant $R$. Said in a different way, once the orientation has been selected so that the D7-brane extends in the five non-compact directions of the background and wraps the three sphere that factors out in \eqref{ds10ds5}, its embedding is described by the two functions ${\phi_1}_{emb}(r)=\phi_1^0$ and $\theta_{emb}(r)=\arcsin(R/r)$, where a dependence on $r$, at most, can be anticipated from the metric. The position $\phi_1^0$ bares no significance, but $R$ is the distance between the extreme D3-brane that generates this particular background and the probe D7-brane that we just embedded on it, and is therefore related to the mass of the quark in the dual gauge theory.

If the extremality condition for the D3-brane is relaxed and a finite horizon is permitted to appear around it, the generated background will not accommodate the flat embedding traced by keeping a constant $R$ in the relationship $\theta_{emb}(r)=\arcsin(R/r)$, and in general the DBI action \eqref{dbi} will have to be written in terms of the functions ${\phi_1}_{emb}(r)$ and $\theta_{emb}(r)$, so that it can be varied with respect to them,  obtaining then the equations they have to satisfy. Since the metric associated to \eqref{ds10ds5} is still diagonal in this case, and its components are functions of $r$ alone, ${\phi_1}_{emb}(r)=\phi_1^0$ persists to be a solution, and dependence on no other coordinate has to be considered for $\theta$. These embeddings and their thermodynamic behavior were numerically constructed and studied in \cite{Kruczenski:2003be,Babington:2003vm,Mateos:2007vn,Mateos:2007yp}, where it was shown that the excess of mass in the D3-brane with respect to the extreme case indeed pulls the D7-brane towards it, bending it to get closer to its horizon than the asymptotic distance $R$. Depending on the value of $R$ in comparison to the radius of the horizon $r_h$, or equivalently, the mass of the quark against the temperature in the dual gauge theory, the D7-brane might reach the horizon or not, leading respectively to the terms black hole embedding or Minkowski embedding, due to the presence of an apparent horizon, or lack thereof, in the worldvolume.

The role of a non-zero value for $B$ in the 10D uplift that we constructed in section \ref{sec:Uplift} for the magnetic brane of \cite{DHoker:2009mmn} is twofold. On the one hand, it confers to some of the metric components a dependence on the $x$ and $y$ coordinates that will potentially reflect on the embedding functions, and on the other, of more consequence, the direction along the $\phi_1$ coordinate is not orthogonal to those spanned by the D7-brane, and therefore the worldvolume cannot remain at constant value for it.

In view of the above we see that the first step to find the way in which the D7-brane is embedded in the ten dimensional background once both, the temperature and the magnetic field are different from zero, is to express the determinant of the metric induced on the worldvolume in terms of $\theta_{emb}$ and ${\phi_1}_{emb}$ as functions of $r, x,$ and $y$. For the calculations ahead, it is more convenient to use $\psi(r,x,y)\equiv \sin\theta_{emb}(r,x,y)$ rather than $\theta$ itself to describe the embedding, and to tidy up the notation, we will refer to ${\phi_1}_{emb}$ as $\varphi$. After implementing this change of variable, the induced metric can be read of \eqref{ds10ds5} once the directions spanned by the D7 are considered, and its determinant is shown in appendix \ref{ApExplEq}. 

The equations for $\psi(r,x,y)$ and $\varphi(r,x,y)$ that result from the variation with respect to them of the DBI action \eqref{dbi} once the determinant of the induced metric has been substituted in it, are lengthy and their general solution presents a technical challenge. The difficulty originates mostly from the fact that, due to the lack of perpendicularity, the way in which the D7-brane most wrap in the $\phi_1$ direction is highly non trivially related to the profile given by $\psi(r,x,y)$, which unavoidably depends on at least one of the coordinates $x$ or $y$.

The focus of our current report is not on this general solution, but rather to present the appearance of Landau levels in the fundamental degrees of freedom when perturbing the equatorial embedding that still proves to be a valid construction in our setting.

Finding other embeddings in the ten dimensional uplift build in section \ref{sec:Uplift} is a task that remains interesting and deserves further exploration. An alternative approach was studied in \cite{Avila:2019pua,Avila:2020ved}, where we found a way to modify the content of the 5D theory so that the $\phi_1$ direction of its uplift was kept perpendicular to those in which the D7-brane could be extended, permitting this object to stay at a fixed value for such coordinate, and also eliminating the necessity for the embedding to depend on $x$ or $y$.

\section{Fundamental Landau levels}\label{SecLL}

Despite the length and complexity of the embedding equations, the direct substitution of $\psi(r,x,y)=0$ and $\varphi(r,x,y)=\varphi_0$, that without trouble can be done using any mathematical software of choice (ours was Mathematica), shows that these functions satisfy them for any real constant $\varphi_0$ regardless of the norm chosen for $A$. As mentioned before, this configuration becomes supersymmetric in the $B\rightarrow 0, T\rightarrow 0$ limit, indicating its stability and encouraging the study of perturbations around it. Of these perturbation, those $\delta\psi$ of the embedding function $\psi+\delta\psi$ are dual to scalar excitations of the fundamental fields in the gauge theory, while those $\delta\varphi$ of $\varphi+\delta\varphi$ correspond to psudoscalar ones \cite{Kobayashi:2006sb,Mateos:2007vn,Myers:2007we}.  As we will explain below, it is consistent to decouple these two channels and keep $\delta\varphi=0$ while studying the behavior of $\delta\psi$, which develops the Landau levels object of the current work. To this end we will study only modes that are constant over the internal 3-cycle, and leave the analysis of those with non-zero R charge for future research.

\subsection{Landau gauge}\label{landaugauge}

Let us first choose Landau gauge $A=B\,x\, dy$ to describe the constant magnetic field $\mathbf{B}=B\,dx\wedge dy$, which corresponds to setting $d\widetilde{\phi}_i=d\phi_i+\frac{2}{\sqrt{3}}B\, x\, dy$ while constructing the uplift.  If we write $\delta\psi(r,t,x,y,z)=\psi_t(t)\psi_x(x)\psi_y(y)\psi_z(z)\psi_r(r)/\psi_r(r_\infty)$, and $\delta\varphi=0$, the equation that came from the variation of \eqref{dbi} with respect to $\delta\varphi(r,t,x,y,z)$ results to be satisfied at all orders, while at leading order in the perturbation the one that resulted from varying with respect to $\delta\psi(r,t,x,y,z)$ reduces to
\be
\begin{split}
& \bigg[3 U V W' \psi_r'+6 W \big(V U' \psi_r'+U V' \psi_r'+U V \psi_r''\big)\\
&+6 V W\left(3 -\frac{\partial_t^2 \psi_t}{U \psi_t}+\frac{\partial_y^2 \psi_y}{V \psi_y}+\frac{\partial_z^2 \psi_z}{W \psi_z} \right) \psi_r\bigg]\psi_x\\
&+W \psi_r \left(6 \partial_x^2\psi_x-8 B^2 x^2 \psi_x\right)=0,
\end{split}
\label{eompsirx}
\ee
which can be separated into
\be
\begin{split}
&\left(3 -\frac{\partial_t^2 \psi_t}{U \psi_t}+\frac{\partial_y^2 \psi_y}{V \psi_y}+\frac{\partial_z^2 \psi_z}{W \psi_z}  \right) V W \psi_r+\frac{1}{2} U V W' \psi_r'\\
&+ W \left(V U' \psi_r'+U V' \psi_r'+U V \psi_r''\right)=2  {\cal{E} } W \psi_r,\end{split}
\label{eompsirL}
\ee
and
\be
\frac{1}{2 }\left[ -\partial_x^2\psi_x+e^2 B^2 x^2 \psi_x\right]={\cal{E}} \psi_x,\label{eompsix}
\ee
where we have used that $U, V,$ and $W$ are functions of $r$ alone and a prime denotes differentiation with respect to this coordinate. We have normalized the perturbation by the value $\psi_r(r_\infty)$, where the limit $r_\infty\rightarrow\infty$ is eventually meant to be taken, to adapt our writing to \cite{Myers:2007we} and therefore be able to use in what follows the holographic dictionary developed in the Appendix A of said work. Notice that dividing the perturbation by $\psi_r(r_\infty)$, or any other constant, does not modify \eqref{eompsirx},  so the calculation above is independent of this normalization. The separation of \eqref{eompsirx} into \eqref{eompsirL} and \eqref{eompsix} has been done in a manner that is amiable for numerical computations and to give \eqref{eompsix} the explicit form of the Schrödinger equation of an harmonic oscillator that describes Landau levels in ordinary Quantum Mechanics for a particle of mass equal to 1, considering that $\hbar$ and the speed of light $c$ have been set to unity throughout our calculation. In the uplift we are working on, the three circles parametrized by the angles $\phi_i$ are treated equally, and they are the geometric dual of the U(1) fiber in the gauge theory, making our writing of the corresponding coupling as $e=\frac{2}{\sqrt{3}}$ an accurate reading of the 1-forms $d\widetilde{\phi}_i=d\phi_i+\frac{2}{\sqrt{3}}B\, x\, dy$. It is interesting to notice the similarity of the latter 1-forms with $dt+\frac{2 n}{L^2}x\, dy$, that bears the NUT charge of the 3+1 planar Taub-NUT black hole in \cite{Cano:2021qzp}, where the study of the excitations of such background leads to the separation of an equation identical in nature to \eqref{eompsix}. 

For $\delta\psi$ to be valid as a perturbation or even for $\psi+\delta\psi\equiv \sin\theta_{emb}$ to be an acceptable description of the D7-brane embedding, $\psi_x$ has to remain bounded for all values of $x$. Demanding that the acceptable solutions to equation \eqref{eompsix} are those that satisfy the latter condition as $x\rightarrow\pm\infty$ leads, in the exact same way as for the harmonic oscillator, to the quantization
\be
{\cal{E}}_n=\left( n+\frac{1}{2}\right)\omega_c, \label{LL1}
\ee
of the separation constant ${\cal{E}}$ in terms of the cyclotron frequency $\omega_c=e\,B$ and the Landau level number, given by the integer $n$. The solution $\psi^{(n)}_x(x)$ to \eqref{eompsix} that corresponds to the value ${\cal{E}}_n$ in \eqref{LL1} is the $n$th normalized Hermite-Gaussian function with argument $\xi=\sqrt{\omega_c}x=\sqrt{eB}x$,
\be
\psi^{(n)}_x(x)=\frac{1}{\sqrt{2^n n!}}\Big(\frac{e B}{\pi}\Big)^{1/4} e^{-\frac{e B x^2}{2}}H_n\Big(\sqrt{e B}x\Big),\label{HF}
\ee
where $H_n$ is the $n$th Hermite polynomial.

\subsection{Symmetric gauge}\label{symmgauge}

We can alternatively chose to describe the same magnetic field using the symmetric gauge $A=\frac{B}{2}(x\, dy\,-\,y\,dx)$, which is implemented by setting $d\widetilde{\phi}_i=d\phi_i+\frac{1}{\sqrt{3}}B(x\, dy\,-\,y\,dx)$ for the construction of the uplift. If in this case we write $\delta\psi(r,t,x,y,z)=\psi_t(t)\psi_z(z)\psi_{xy}(x,y)\psi_r(r)/\psi_r(r_\infty)$ while keeping $\delta\varphi=0$ as before, the only term that appears at leading perturbative order on the equation coming from the variation of  \eqref{dbi} with respect to $\delta\psi$ separates as it did in section \ref{landaugauge}, with the only difference that now we have
\be
\begin{split}
&\left(3 -\frac{\partial_t^2 \psi_t}{U \psi_t}+\frac{\partial_z^2 \psi_z}{W \psi_z}  \right) V W \psi_r+\frac{1}{2} U V W' \psi_r'\\
&+ W \left(V U' \psi_r'+U V' \psi_r'+U V \psi_r''\right)=2  {\cal{E} } W \psi_r,\end{split}
\label{eompsirs}
\ee
and
\be
\frac{1}{2}\left[ -\partial_x^2\psi_{xy}-\partial_y^2\psi_{xy}+\frac{1}{4}e^2 B^2( x^2+y^2) \psi_{xy}\right]={\cal{E}} \psi_{xy},\label{eompsixy}
\ee
instead of \eqref{eompsirL} and \eqref{eompsix}.

In this gauge the equation resulting from varying with respect to $\delta\varphi$ is not satisfied automatically, and at leading perturbative order reduces to
\be
y\partial_x\psi_{xy}-x\partial_y\psi_{xy}=0,\label{nullam}
\ee
which left hand side is interestingly proportional to the canonical angular momentum operator
\be
L_z\equiv -i(x\partial_y-y\partial_x).
\ee

We therefore see that any function $\psi_{xy}$ that simultaneously solves \eqref{eompsixy} and \eqref{nullam} is also a solution to the equation
\be
\begin{split}
\frac{1}{2}&\big[ -\partial_x^2\psi_{xy}-\partial_y^2\psi_{xy}-e\,B\,L_z\psi_{xy}\\
&+\frac{1}{4}e^2 B^2( x^2+y^2) \psi_{xy}\big]={\cal{E}}\, \psi_{xy},
\end{split}
\label{eomLLsym}
\ee
that describes Landau quantization in Quantum Mechanics when using the symmetric gauge.

The quantization condition \eqref{LL1} is now identically recovered from the requirement for $\psi_{xy}$ to be well behaved, in the same way it would happen for the states of vanishing canonical angular momentum of a two dimensional isotropic harmonic oscillator with frequency $\omega_c/2$. 

From the analysis above we can claim that through our construction in the symmetric gauge, all Landau levels with vanishing canonical angular momentum, and only those, are recovered. This does not mean that there is no rotation in the states we are able to reproduce, since the kinetic angular momentum is given by the operator
\be
{\mathcal{L}}_z\equiv x(-i\partial_y -e\,A_y)-y(-i\partial_x -e\,A_x),
\ee
which in our setting, and given the $L_z\psi_{xy}=0$ restriction, reduces to
\be
{\mathcal{L}}_z=-\frac{e\,B}{2}(x^2+y^2),
\ee
when acting over our set of solutions $\psi_{xy}$.

In ordinary Quantum Mechanics, where $\psi_{xy}$ is the wave function, the expectation value $\langle x^2+y^2 \rangle$ in the $n$th Landau level with $l_z=0$ is given by $\frac{2}{e\,B}(n+1)$. If this consideration was valid in our setting, it would lead us to associate a kinetic angular momentum equal to $n+1$, in $\hbar$ units, and a cyclotron frequency given again by $\omega_c=e\,B$, to the solutions we found in this section.

\section{Radial profile of $\delta\psi$}\label{solution}

We now proceed to numerically calculate the solutions of \eqref{eompsirL} and  \eqref{eompsirs} to, as a first step, demonstrate that they exist in a consistent manner with the perturbative approach.  Working in the rest frame of the plasma both \eqref{eompsirL} and  \eqref{eompsirs} reduce, after Fourier transforming, to
\be
\begin{split}
&\left(3 +\frac{\omega^2}{U} \right) V W \psi_r+\frac{1}{2} U V W' \psi_r'\\
&+ W \left(V U' \psi_r'+U V' \psi_r'+U V \psi_r''\right)=2 {\cal{E} } W \psi_r,\end{split}
\label{eompsir}
\ee
and since \eqref{LL1} dictates the value of the separation constant ${\cal{E} }$ to be $(n+1/2)e\, b$ for either $\psi_x$ or $\psi_{xy}$, we notice that the analysis of the radial function $\psi_r$ in the rest frame of the plasma is identical in either of the two gauges we have explored.

As explained in Appendix A of \cite{Arean:2016het}, the elements of the family of backgrounds that we are interested on can be constructed by writing the metric functions $U(r)$, $V(r)$, and $W(r)$ as the series
\be
\begin{split}
&U(r)=6 r_h (r-r_h)+\sum _{\text{i}=2}^\infty U_i(r-r_h)^i,\\
&V(r)=\sum _{\text{i}=0}^\infty V_i(r-r_h)^i,\\
&W(r)=3\,{r_h}^2 (r-r_h)^0 +\sum _{\text{i}=1}^\infty W_i(r-r_h)^i,
\end{split}\label{seriesUVW}
\ee
which introduce a horizon located at $r=r_h$, and provide sufficient initial conditions near this location to integrate equations \eqref{eqback} towards larger values of $r$. If we work at a given fixed temperature $T=U'(r_h)/4\pi=\frac{3}{2\pi}r_h$, the only free coefficient that is not determined by the equations \eqref{eqback} is $V_0$, and therefore its value can be used along with that of $B$ to construct different backgrounds.

Any solution obtained in this way presents the limiting behavior
\be
\lim_{r\rightarrow\infty}(r)\rightarrow r^2, \lim_{r\rightarrow\infty}V(r)\rightarrow V_\infty r^2,\lim_{r\rightarrow\infty}W(r)\rightarrow W_\infty r^2,
\ee
with $V_\infty$ and $W_\infty$ constants that depend on the values of $B$ and $V_0$ in \eqref{seriesUVW}.

To bring the resulting metric to one that is asymptotically AdS, we use the invariance shown by equations \eqref{eqback} when scaling either $W(r)$ or simultaneously $V(r)$ and $B$, and write the metric functions
\be
U(r),\quad \tilde{V}(r)\equiv V(r)/V_\infty, \quad\mbox{and}\quad \tilde{W}(r)\equiv W(r)/W_\infty,
\ee
which solve the same equations for $B\rightarrow b=B/V_\infty$ and feature the desired asymptotic behavior, without changing the temperature $T=U'(r_h)/4\pi=\frac{3}{2\pi}r_h$. In what follows we will assume this scaling has already been performed, including on the coefficients in \eqref{seriesUVW}, and dispense with the tilde. 

An important feature is that regardless of the individual values of $V_0$ and $B$, the same ratio $B/V_0$ will lead to the same solution after scaling, hence we can fix $V_0$ in the construction process and use $B$ to sweep the whole family. More details about these backgrounds as a set can be found in \cite{Arean:2016het} and \cite{Martinez-y-Romero:2017awl}, but one more fact we want to mention for the sake of clarity below is that for any given values of $b$ and  $T=\frac{3}{2\pi}r_h$, there is a unique solution to \eqref{eqback} that is asymptotically AdS and has a near horizon geometry described by \eqref{seriesUVW}. From this we see that, at fixed temperature, we can label the elements of the family using $b$, which indeed runs from zero to infinity. 

For large $r$ the metric functions can be approximated as
\be
\begin{split}
    &U(r)=r^{2}+U_{1}r+\frac{U_{1}^{2}}{4}+\frac{1}{r^{2}}\left(U_{-2}-\frac{2}{3}b^{2}\log{r}\right)\\
    &+U_1\frac{1}{r^{3}}\left(-U_{-2}-\frac{1}{3}b^2+\frac{2}{3}b^{2}\log{r}\right)+\mathcal{O}\left(\frac{1}{r^{4}}\right),\\
    &V(r)=r^{2}+U_{1}r+\frac{U_{1}^{2}}{4}+\frac{1}{r^{2}}\left(V_{-2}+\frac{1}{3}b^{2}\log{r}\right)\\
    &+U_1\frac{1}{r^{3}}\left(-V_{-2}+\frac{1}{6}b^2-\frac{1}{3}b^{2}\log{r}\right)+\mathcal{O}\left(\frac{1}{r^{4}}\right),\\
    &W(r)=r^{2}+U_{1}r+\frac{U_{1}^{2}}{4}+\frac{1}{r^{2}}\left(-2V_{-2}-\frac{2}{3}b^{2}\log{r}\right)\\
    &+U_1\frac{1}{r^{3}}\left(2V_{-2}-\frac{1}{3}b^2+\frac{2}{3}b^{2}\log{r}\right)+\mathcal{O}\left(\frac{1}{r^{4}}\right),
\end{split}\label{BseriesUVW}
\ee
where the free parameters $U_1, U_{-2},$ and $V_{-2}$ can be related to those in the near horizon expansion by analyzing the numerical solutions constructed using specific values for the latter.

Using \eqref{seriesUVW} in \eqref{eompsir} we determine that near the horizon $\psi_r(r)$ behaves like $(r-r_h)^{i\alpha}$ for $\alpha=\pm \frac{\omega}{4\pi T}=\pm \frac{\omega}{6 r_h}$, of which we choose the negative sign to impose the ingoing wave boundary condition and develop the expansion
\be
\begin{split}
&\psi^{(\omega,n)}_r(r)\simeq(r-r_h)^{-i\frac{\omega}{6 r_h}} \psi_r^{(0)}\\
&\left[1+C_{(r,\omega,n)}^{(1)}(r-r_h)+C_{(r,\omega,n)}^{(2)}(r-r_h)^2+{\mathcal{O}}(r-r_h)^3\right],
\end{split}
\label{seriespsi}
\ee
where $\psi_r^{(0)}$ is a free parameter that can be adjusted to keep the perturbative approximation valid, and it appears as a global factor due to the linear character of \eqref{eompsir}, while
{\footnotesize
\be
\begin{split}
    &C_{(r,\omega,n)}^{(1)}=\frac{1}{108 {r_h}^2 V(r_h)^2 (3 {r_h} - i\omega)}\big\{b^2\omega (5\omega - 3 i r_h) \\
    &- 6 V(r_h)\left[-18 (n+1/2)e\, b {r_h}^2  + V(r_h)\left (27 {r_h}^2 - 15 i {r_h}\omega + \omega^2 \right) \right]\big\}\\
    &C_{(r,\omega,n)}^{(2)}=\frac{1}{23328 {r_h}^4 V(r_h)^4 \left(18
   {r_h}^2-9 i {r_h} \omega -\omega ^2\right)}\\
   &\Big\{b^4 \omega  \left(-252 {r_h}^2
   \omega -468 i {r_h}^3+35 i {r_h} \omega ^2+25 \omega ^3\right)+\\
   & 12 b^2 V(r_h) \Big[18 (n+1/2)e\, b {r_h}^2 
   \left(36 {r_h}^2-12 i {r_h} \omega +5 \omega ^2\right)\\
   &+V(r_h) \left(-459 {r_h}^2 \omega ^2-288 i {r_h}^3
   \omega +972 {r_h}^4+97 i {r_h} \omega ^3-5 \omega ^4\right)\Big]+\\
   &36 V(r_h)^2 \Big[324 (n+1/2)^2 e^2\, b^2 {r_h}^4 -36 (n+1/2)e\, b {r_h}^2 V(r_h)\\
   &\left(99  {r_h}^2-24 i {r_h} \omega +\omega ^2\right)\\
   &+V(r_h)^2 \left(-126 {r_h}^2 \omega ^2-2268 i {r_h}^3 \omega
   +2673 {r_h}^4-45 i {r_h} \omega ^3+\omega ^4\right)\Big]\Big\}.
\end{split}
\label{coefseriespsi}
\ee }

A non trivial consistency check for the quantization condition \eqref{LL1} is provided by \eqref{coefseriespsi}, since only if the separation constant ${\mathcal{E}}_n$ is proportional to $b$, as \eqref{LL1} demands, then the near horizon behavior of $\psi_r$ dictated by \eqref{coefseriespsi} depends on the ratio $b/V(r_h)=B/V_0$ and not on $b$ and $V(r_h)$ separately.

We can now use \eqref{seriespsi} to evaluate $\psi_r$ and its derivative with respect to $r$ at $r=r_h+\epsilon$, where the value of $\epsilon$ is selected to achieve the desire accuracy when utilizing the results of such evaluation as initial conditions to numerically integrate towards large $r$.

The substitution of \eqref{BseriesUVW} in \eqref{eompsir} shows that as $r\rightarrow\infty$, $\psi_r(r)$ can be approximated by
\be
\begin{split}
&\psi^{(\omega,n)}_r(r)\simeq {\psi_r}^{(-1)}\bigg[\frac{1}{r}-\frac{U_1}{2r^2}+\left(\omega^2-2 (n+1/2)e\, b\right)\frac{\log{r}}{2 r^3}\\
&-3 U_1 \left(\omega^2-2 (n+1/2)e\, b\right)\frac{\log{r}}{4 r^4}+\\
& U_1 \left(\omega^2-2 (n+1/2)e\, b+{U_1}^2\right)\frac{1}{4 r^4}\bigg]\\
&+{\psi_r}^{(-3)}\left[\frac{1}{r^3}-6 U_1 \frac{1}{4 r^4}\right]+\mathcal{O}\left(\frac{1}{r^{5}}\right),
\end{split}
\label{Bseriespsi}
\ee
where ${\psi_r}^{(-1)}$ and ${\psi_r}^{(-3)}$ are expansion coefficients, and we see that there is no imminent divergent behavior that would invalidate the perturbative approach. It will be important below that, as the authors in \cite{Myers:2007we} mention, the dictionary that they develop in their appendix A applies to fields with general dependence in the gauge theory directions, and therefore we see that the source term for the scalar operator dual to $\delta\psi$ in our case is given by
\be
\mu({\mathbf{x}})=\frac{1}{2}\sqrt{\lambda}T{\psi_r}^{(-1)}\psi_{{\mathbf{x}}}({\mathbf{x}}),\label{source}
\ee
where $\psi_{{\mathbf{x}}}({\mathbf{x}})$ denotes either $\psi_{x}(x)$ or $\psi_{xy}(x,y)$.

\begin{figure}
\begin{subfigure} {0.4\textwidth}
\centering
\includegraphics[scale=.35]{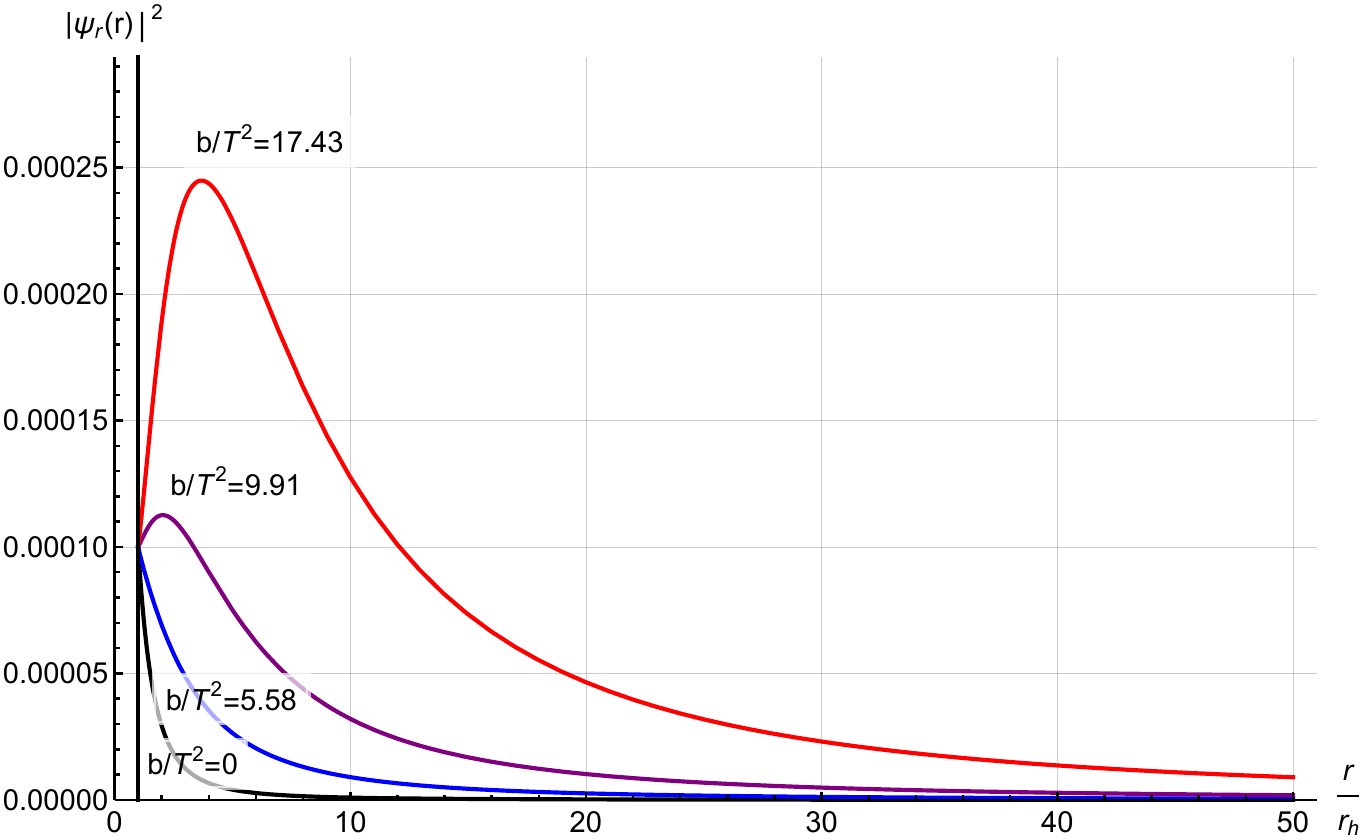} \nonumber
\caption{$n=5, \omega/\pi T=2$}
\end{subfigure}
\begin{subfigure} {0.4\textwidth}
\centering
\includegraphics[scale=.33]{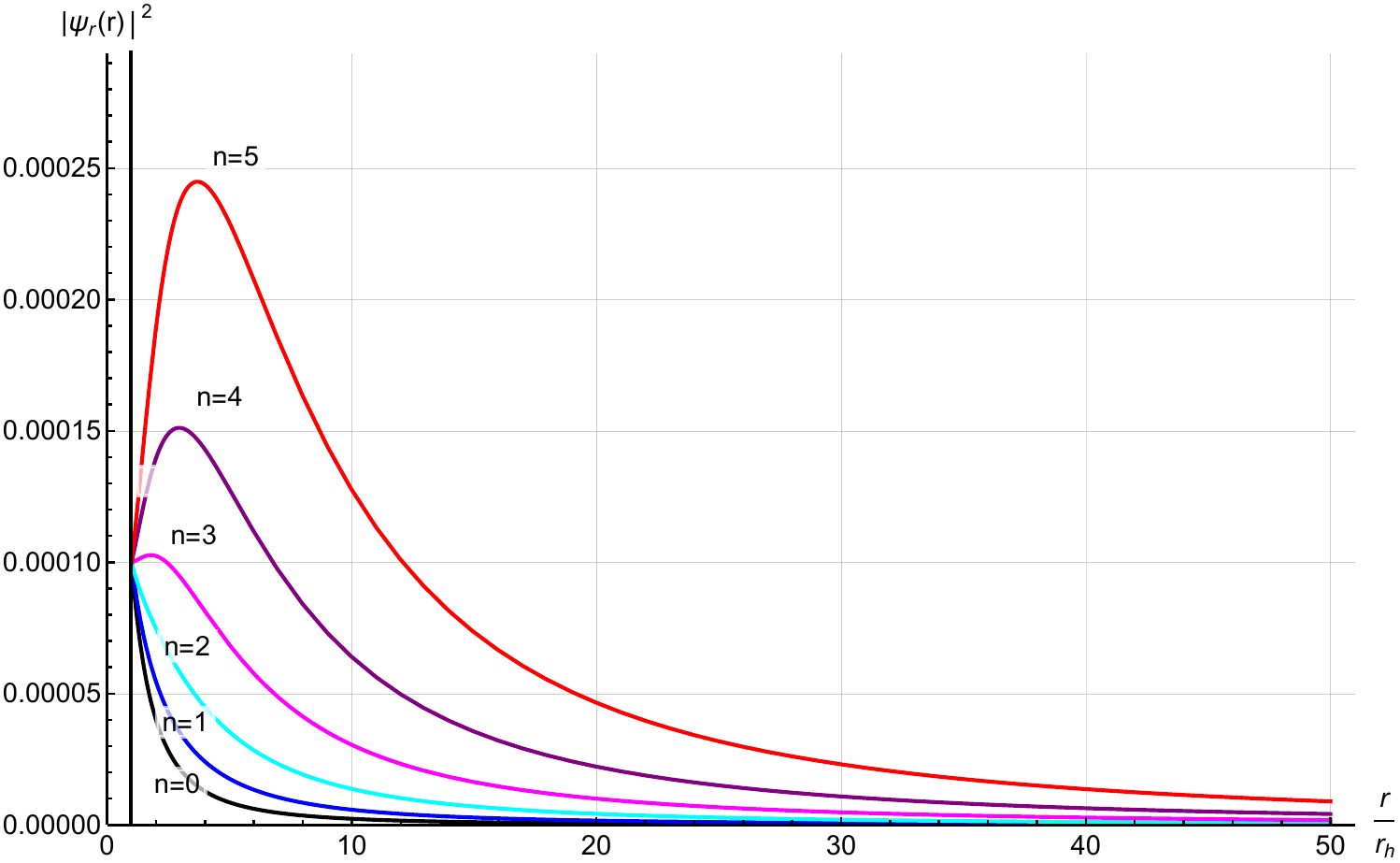} 
\caption{$b/T^2=17.43, \omega/\pi T=2$}
\end{subfigure}
\begin{subfigure} {0.4\textwidth}
\centering
\includegraphics[scale=.33]{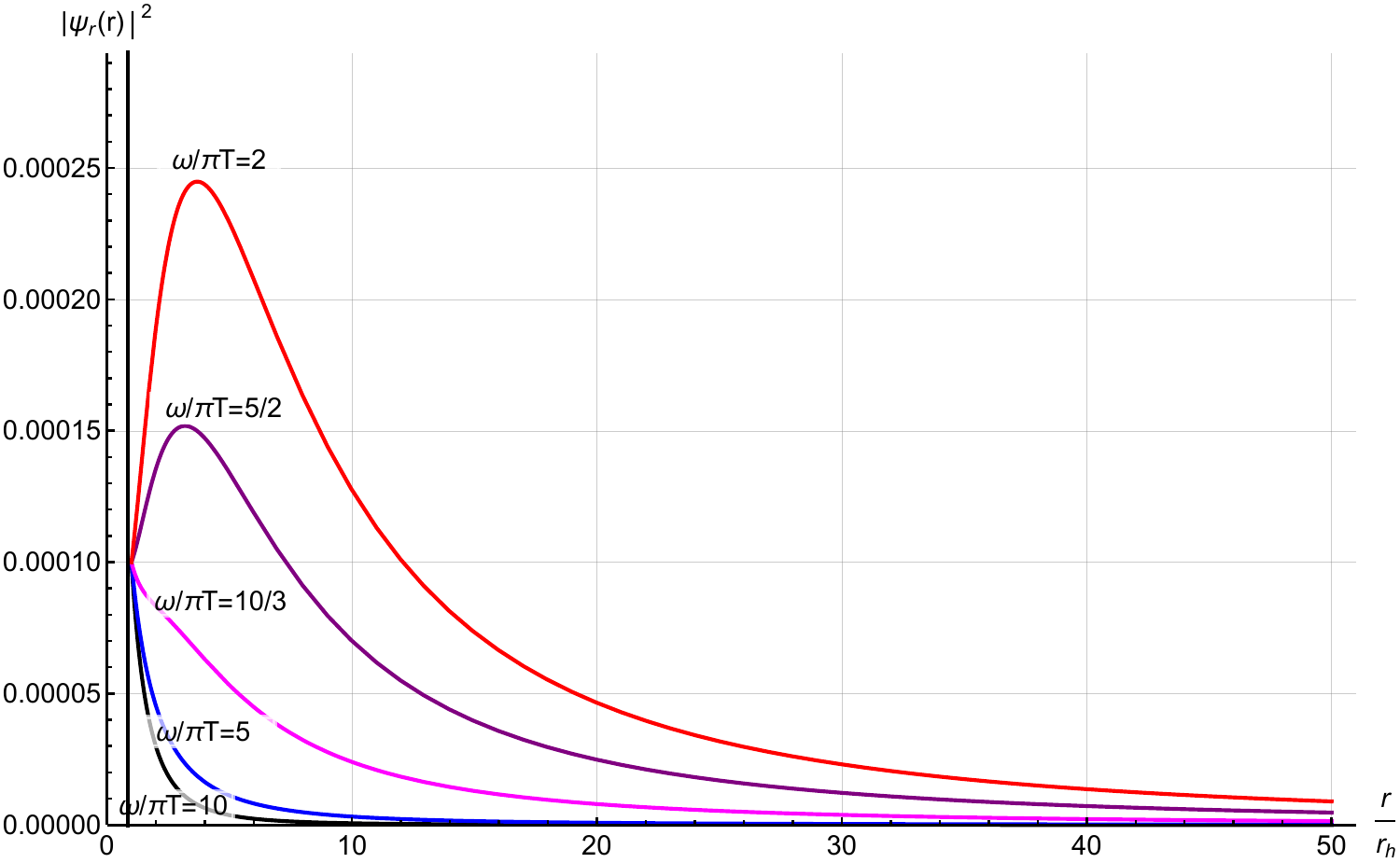} \nonumber
\caption{$b/T^2=17.43, n=5$}
\end{subfigure}%
    \caption{Plots of the square of the radial profile $\psi_r(r)$ of the perturbation $\delta\psi$. In each of the graphics two of three quantities $\{ b/T^2, n, \omega/\pi T\}$ have been kept fixed at the value indicated below them, while the third has been varied as the label of each curve indicates. The horizon appears as a vertical line because of the scale, which is set by the size of $\psi_r^{(0)}/r_h $.}
    \label{rProfile}
\end{figure}

Before using the results in this section to extract the quasinormal modes of the scalar operator,  let us present in Fig. \ref{rProfile} the plots of the square of the complex radial profiles $\vert\psi_r(r)\vert^2$ numerically obtained from a starting value $\psi_r^{(0)}=1/100$. We have arrange three panels to exhibit how changing the values of each of the quantities $b/T^2, n,$ and $\omega/\pi T$, affect the behavior of $\vert\psi_r(r)\vert^2$ when the other two are kept fixed. 

The intention behind Fig. \ref{rProfile} is to make visually clear that the real and imaginary parts of the profile function $\psi_r(r)$ both remain bounded across the entire domain and hence can be kept in the perturbative regime by using a small enough value for $\psi_r^{(0)}$, validating the applicability of our calculations.

\section{Quasinormal modes}\label{QNM}

The quasinormal modes of the scalar excitations of the fundamental fields that we have studied are those for which the source $\mu$ vanishes, and therefore their dual is given by the solutions $\delta\psi$ with a normalizable radial profile characterize by $\psi^{(-1)}=0$ in the asymptotic behavior \eqref{Bseriespsi}.

Due to the numeric nature of our construction, we resourced to a two-dimensional shooting method, briefly explained in appendix \ref{shottingmeth}, to find the complex value of $\omega$ that for a given ${\mathcal{E}}_n=(n+1/2)e\, b $ leads to the vanishing of  $\psi^{(-1)}$ in the boundary expansion of the solution that is ingoing at the horizon. The procedure above shows that, for quasinormal modes, there is a relationship between the separation constant ${\mathcal{E}}_n$ and the frequency $\omega$, which up to this point had been treated as independent parameters.

\begin{figure}
\begin{subfigure} {0.5\textwidth}
\caption{Re[$\omega$]}
\includegraphics[scale=.5]{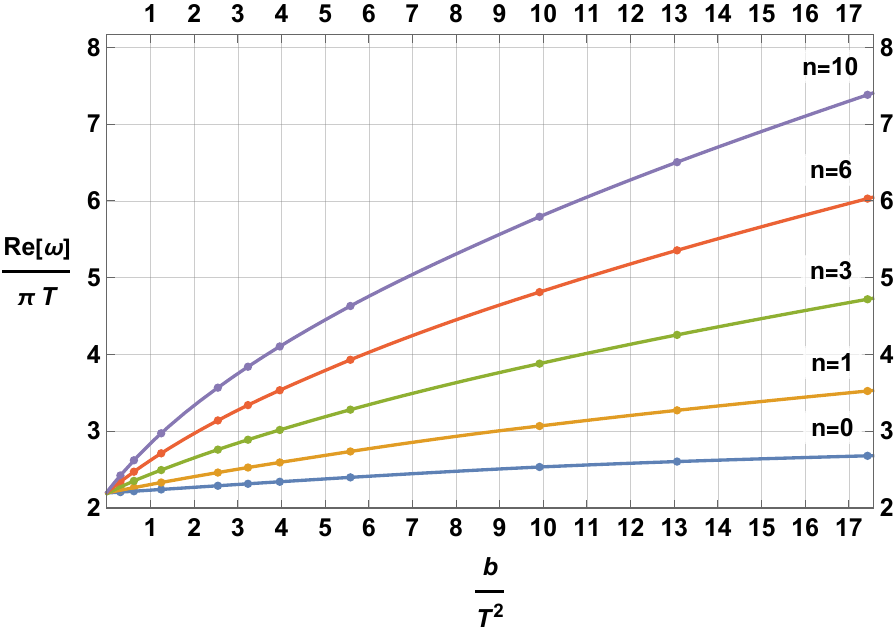} \nonumber
\end{subfigure}
\begin{subfigure} {0.5\textwidth}
\caption{Im[$\omega$]}
\includegraphics[scale=.58]{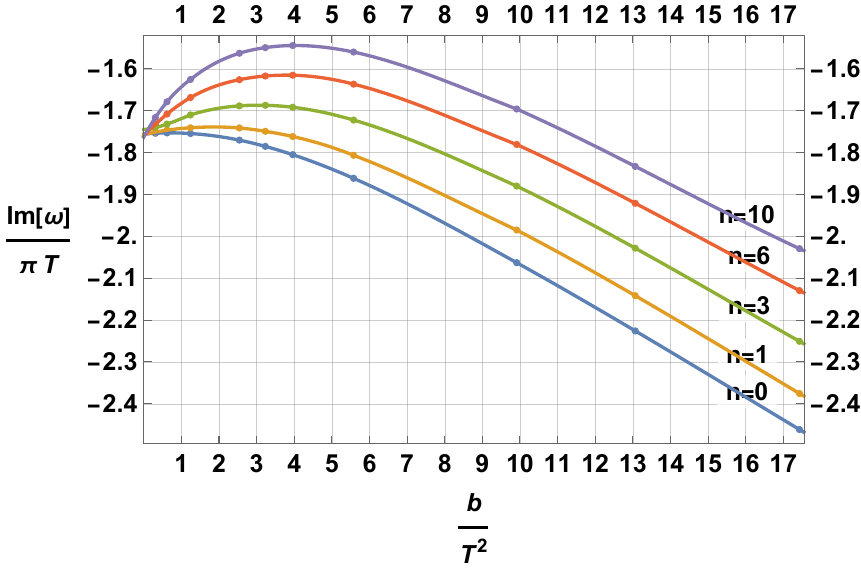} 
\end{subfigure}%
    \caption{Plots of the  real and imaginary parts of the lowest quasinormal frequency as a function of the intensity of the magnetic field in temperature square units. We use $\omega/\pi T$ as a dimensionless quantity to be able to do direct comparisons with previous work such as \cite{Hoyos-Badajoz:2006dzi}, where the ingoing frequency is $\omega/4$ unlike our $\alpha=\omega/4\pi T$.}
    \label{Omvsb}
\end{figure}

Being the more stable, we focus on the lowest quasinormal frequency, of which in Figs. \ref{Omvsb} we present the real an imaginary parts as a function of $b/T^2$ for some representative Landau levels. We immediately see that quasinormal modes of higher Landau levels are more stable since the magnitude of the imaginary component of their frequency decreases with $n$, coinciding with the expectation that a non-zero kinetic energy with respect to the plasma rest frame should have a stabilizing effect.

A more precise reading can be obtained form the energy and the width of the unstable states dual to the quasinormal modes. The energy of such modes can be computed as $E=\omega_{\mathrm{Re}}\equiv\sqrt{{\mathrm{Re}}[\omega^2]}$, since the value $E_0\equiv m_0$, that results of this expression at zero spatial momentum, corresponds to the real part of the pole in the propagator of the scalar operators we are studying. The corresponding width $\Gamma$ can be obtain from the equality $m_0\Gamma=2{\mathrm{Re}}[\omega]{\mathrm{Im}}[\omega]$ that follows from the relativistic Breit-Wigner formula.

In the absence of thermal or destabilizing effects, the spectrum of Landau energies follows the dispersion relation
\be
E=\sqrt{{m_0}^2+e\, b\,(n+1/2)}. \label{exactDispersion}
\ee

\begin{figure}
\includegraphics[scale=.37]{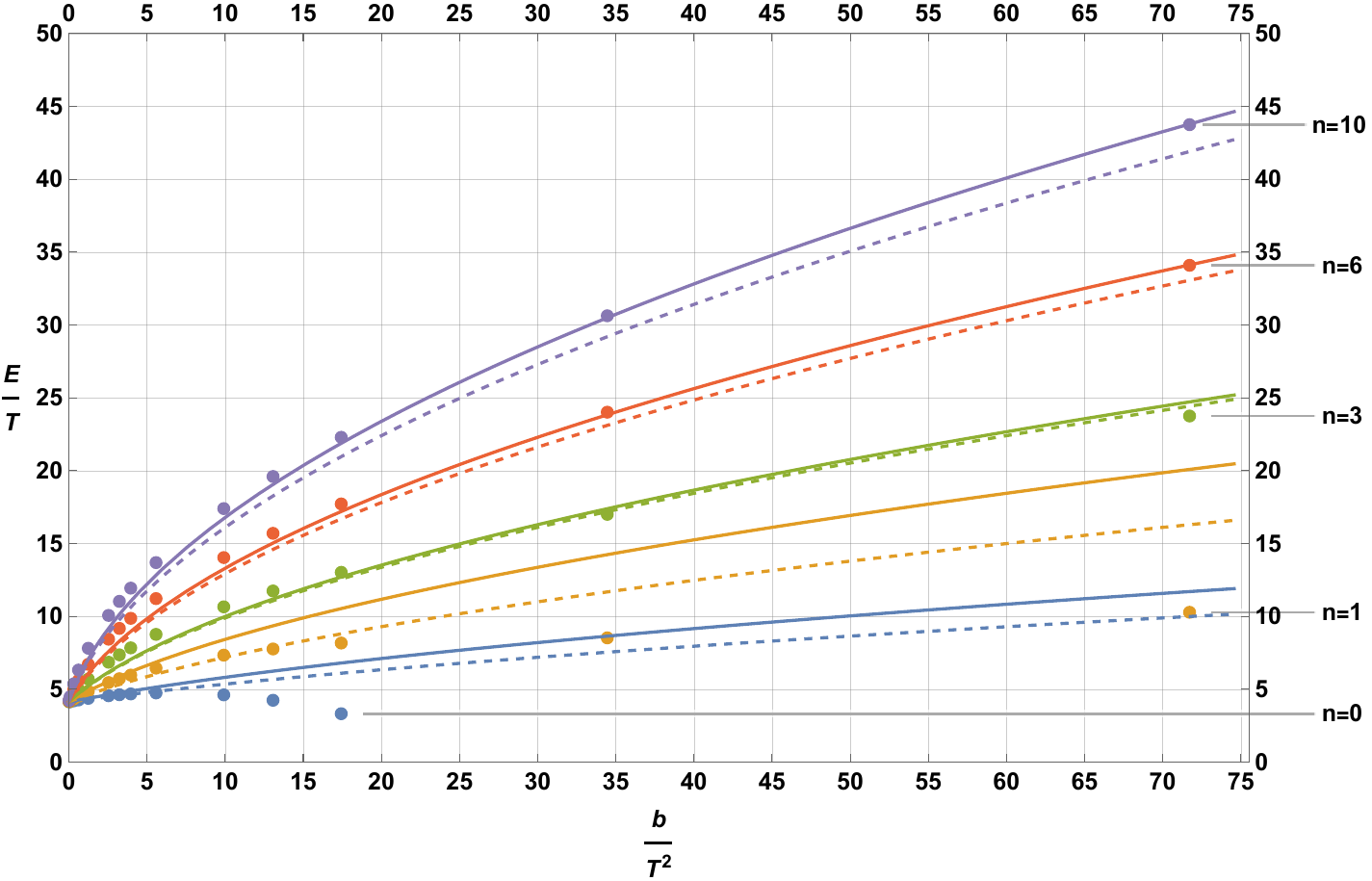}
\caption{Energy as a function of the magnetic field in thermal units. The dots indicate numerically obtained data, the solid lines are traced by the dispersion relations $E/T=\sqrt{{\tilde{m_0}}^2+\tilde{e} (b/T^2)(n+1/2)}$ with the values for $\tilde{m_0}$ and $\tilde{e}$ that best fit each level, and the dashed lines show the prediction for stable Landau levels given by same relations but with our actual parameters $m_0$ and $e$.}
\label{EvslargeB}
\end{figure}

In Fig. \ref{EvslargeB} we have plotted our numerical results for $E/T$ as a function of $b/T^2$ for the representative levels $n=\left\lbrace 0,1,3,6,10\right\rbrace$, along with the curves traced by dispersion relations $E/T=\sqrt{{\tilde{m_0}}^2+\tilde{e} (b/T^2)(n+1/2)}$ for the values of $\tilde{m_0}$ and $\tilde{e}$ that best fit each of them.

\begin{figure}
\includegraphics[scale=.41]{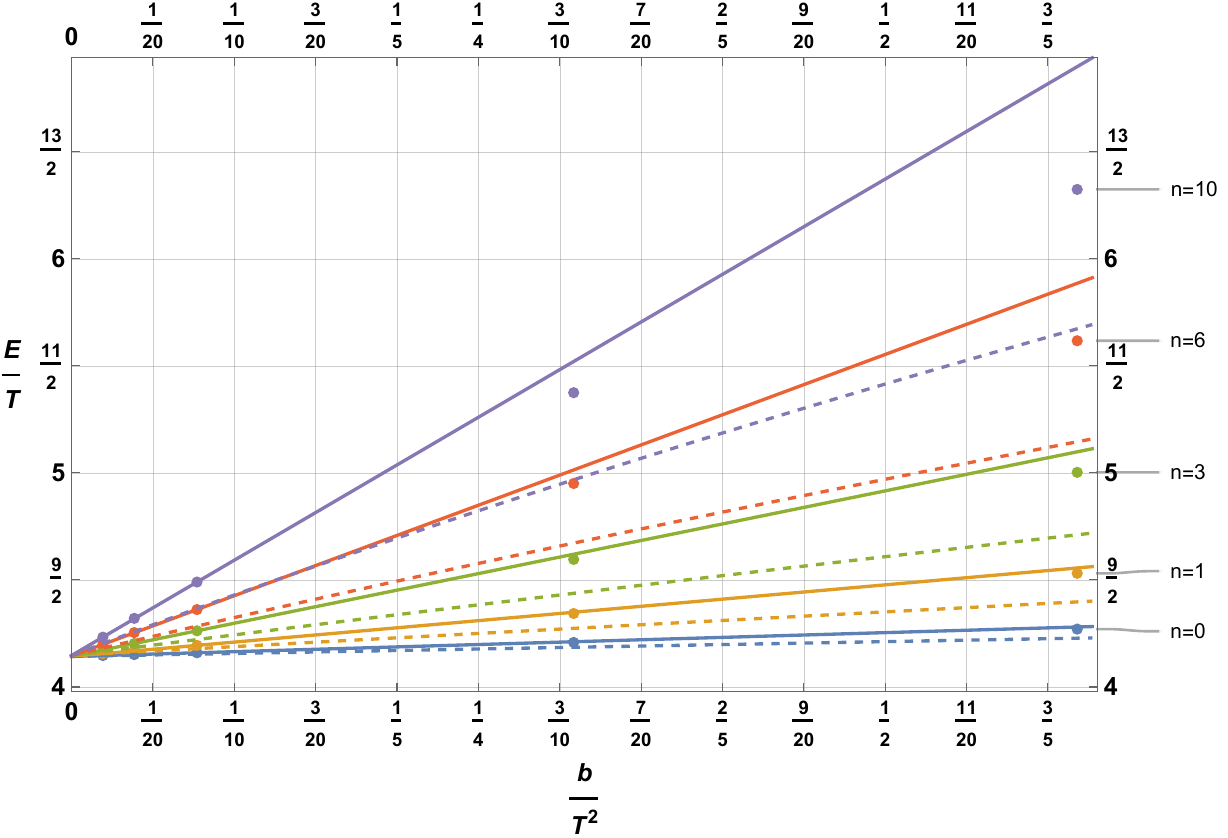}
\caption{Energy as a function of the magnetic field in thermal units. The dots indicate numerically obtained data, the solid lines are traced by the linear relations that best fit each level, and the dashed plots show the linear approximation of the prediction for stable Landau levels corresponding to our actual parameters $m_0$ and $e$.}
\label{EvssmallB}
\end{figure}

The detail of the region $b/T^2\ll 1$ is depicted in Fig.  \ref{EvssmallB},  where we observe, and fit, a linear behavior of the type $E/T\sim \tilde{m_0}/T+ \frac{1}{2} \frac{\tilde{e} (n+1/2)}{\tilde{m_0}/T} B/T^2$, expected from the expansion of \eqref{exactDispersion} around $b=0$. As it should be, the value $E_0/T$ that $E/T$ takes at $b/T^2=0$ is the same for all levels and coincides with the results previously reported \cite{Hoyos-Badajoz:2006dzi} in the absence of magnetic field. This last observation permits us to extract $m_0\equiv E_0$, and use this value to include in both Figs. \ref{EvslargeB} and \ref{EvssmallB} the plots of the dispersion relations $E/T=\sqrt{{m_0}^2/T^2+e\, b/T^2\,(n+1/2)}$ that fully stable Landau levels would satisfy. It is noticeable that despite the instability inherent to the quasinormal modes, our numerical results are not far from the expected energies for stable Landau levels.

\begin{figure}
\includegraphics[scale=.38]{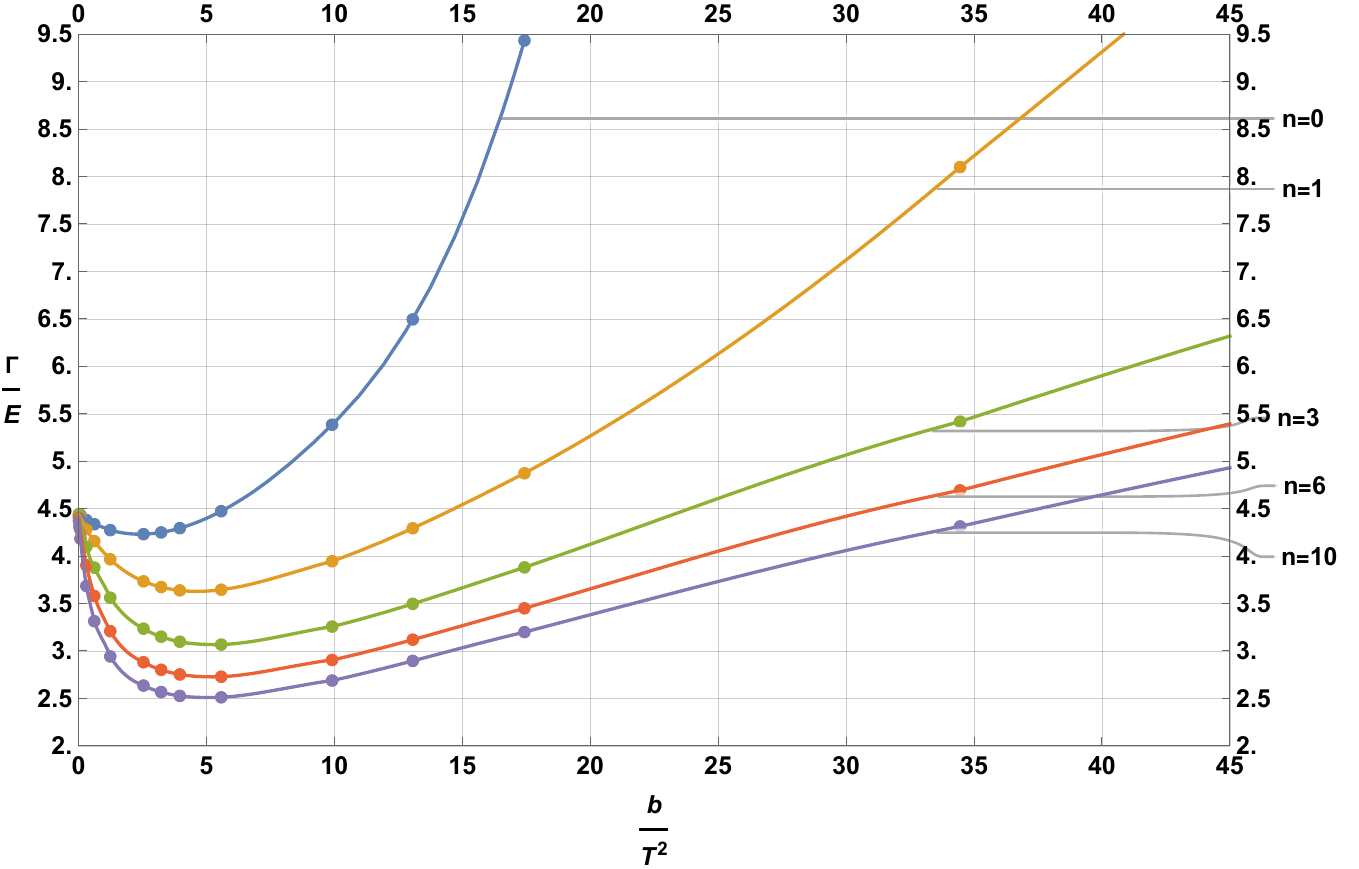}
\caption{Width over energy of the states as a function of the magnetic field in thermal units. The dots indicate numerically obtained data.}
\label{gammaoverE}
\end{figure}

To understand why there are regions in which our states have a larger deviation from the expected Landau levels, in Fig. \ref{gammaoverE} we plot $\Gamma/E$ as a function of $b/T^2$ for each $n$. From this we see that, aside differences to be expected from thermal effects, our results satisfy the dispersion relation of Landau levels in regions where the width of the quisinormal mode is small in comparison to its energy.

It is also interesting to notice from Fig. \ref{gammaoverE} that the effect of the magnetic field is stabilizing for intensities up to a critical value $\frac{b}{T^2}\vert_c (n)$, that increases with the level number $n$, after which the phenomenon is reversed. This is in qualitative agreement with the results in \cite{Ayala:2014yla}, where using resummation techniques, a phenomenon of inverse magnetic catalysis is found for small intensities of the magnetic field, that turns into catalysis when the intensity surpasses a critical value.

The behavior presented so far in this section provides enough evidence for the evolution of the reported quasinormal modes to be considered as the high temperature dissociation of the low temperature stable Landau levels.

\section{Discussion}\label{discu}

As we saw in section \ref{landaugauge}, the behavior over the spatial directions of the excitations of the fundamental scalar field studied here is govern by an equation that is identical in appearance to the one describing Landau quantization in a translational invariant gauge. In section \ref{symmgauge} we were also able to write a system of two equations that as acceptable solutions has the functions that describe Landau levels with vanishing canonical angular momentum in a rotational symmetric gauge, for which we provided the expression to obtain their non-zero kinetic angular momentum. It is interesting to notice that in \cite{Ammon:2017ded} the modes that behaved like Landau levels were also only found in the helicity-0 sector, posing the question about whether this projection in a strongly coupled plasma is more relevant than it would seem at a first glance.

To complete the analysis of the scalar perturbations object of our work, we turned to the study of the quasinormal modes of the system by determining the frequencies leading to normalizable perturbations of the D7-brane equatorial embedding. As graphically displayed in Fig. \ref{EvslargeB}, our numerical results show a behavior for the energy $E\equiv \sqrt{{\mathrm{Re}}[\omega^2]}$ of the states dual to such modes that is remarkably close to the one followed by the energy of fully stable Landau levels as a function of the intensity of the magnetic field. As it should be, at $b=0$ all levels converge to the same value for their lowest quasinormal frequency, which coincides with the one found in \cite{Hoyos-Badajoz:2006dzi}.

The approximated coincidence alluded above provides evidence that the evolution of the states that we found describes the decay of Landau levels that became unstable due to thermal effects. To provide a measure of the relaxation time, we studied the width $\Gamma$ of the modes, and notice that the magnetic field has an stabilizing effect up to a certain intensity, above which, the phenomenon is reversed and $\Gamma$ begins to grow with $b$. This latter observation is in agreement with the resummation calculations showing the existence of inverse magnetic catalysis up to a certain intensity of the magnetic field, above which magnetic catalysis is observed.

There are at least two other topics we believe deserve further investigation. One is the possibility of adapting our construction to recover a fuller version of Schrödinger equation when minimally coupled to a U(1) potential.  A first step in this direction would be to construct the uplift of the charged magnetic brane \cite{DHoker:2009ixq}, since the inclusion of an electric component of the field may lead to interesting phenomena \cite{Edery:2018nfs}.

The other aspect we believe is worth additional consideration is the pursue of embeddings other than the one located at $\psi=0$ and constant $\varphi$, accompanied by the study of their quasinormal modes. In this last direction, we have already found complex solutions to our exact embedding equations, which analyzed from the point of view in \cite{Koerber:2010bx,Abt:2019tas,Grana:2020hyu,Nakas:2020hyo} could lead to interesting results.

\section{Acknowledgments}

We acknowledge partial financial support from PAPIIT IN113618, UNAM.

\clearpage

\appendix

\section{Determinant of the induced metric on the D7-brane for a general gauge}\label{ApExplEq}

If a general $A_\mu$ is considered in $d\widetilde{\phi}_i\equiv d\phi_i+\frac{2}{\sqrt{3}}A_\mu dx^\mu$ when constructing \eqref{ds10ds5}, the number of off diagonal elements in the metric becomes unnecessarily large for our target calculation. We can conveniently use $A=B[\alpha y\,dx+(1+\alpha)x\,dy]$, which results into $F_{GT}=B\,dx\wedge dy$ regardless of the gauge set by the parameter $\alpha$, that includes the particular cases of the translationally invariant for $\alpha=0$, and the symmetric one for $\alpha=-1/2$.

Once $d\widetilde{\phi}_i= d\phi_i+\frac{2}{\sqrt{3}}B[\alpha y\,dx+(1+\alpha)x\,dy]$ has been substituted in \eqref{ds10ds5}, we compute the metric ${g_{D7}}_{I'J'}=\frac{\partial I'}{\partial I}\frac{\partial J'}{\partial J}g_{IJ}$ induced on the D7-brane by using as its coordinate those of the background corresponding to $(t,r,x,y,z,\vartheta,\phi_2,\phi_3)$, and writing the embedding functions $\theta=\arcsin[\psi(r',t',x',y')]$ and $\phi_1=\varphi(r',t',x',y')$. The determinant of the metric $g_{D7}$ that results of this procedure reduces to
\be
\begin{split}
&\frac{1}{3} \sin^2{\vartheta}\cos^2{\vartheta}\, W \left(\psi ^2-1\right)^2\\
&\times\Big[3 \Big(U (\partial_r{\varphi})^2 \psi ^4+\big\{1-U (\partial_r{\varphi})^2\big\} \psi ^2-U (\partial_r{\psi})^2-1\Big) V^2\\
&-\psi ^2 \Big(\big\{4 B^2 y^2 \alpha ^2+4 B y \sqrt{3} {\partial_x{\varphi}} \alpha +3 (\partial_x{\varphi})^2\big\} (\partial_y{\psi})^2\\
&+\big\{4 B^2 x^2 (\alpha +1)^2+4 B x \sqrt{3} {\partial_y{\varphi}} (\alpha +1)+3 (\partial_y{\varphi})^2\big\} (\partial_x{\psi})^2\\
&-2{\partial_x{\psi}} {\partial_y{\psi}} \big\{ {\partial_y{\varphi}} \big[2 B y \alpha  \sqrt{3}+3 {\partial_x{\varphi}}\big]\\
&+2 B x (\alpha +1) \big[2 B y \alpha +\sqrt{3} {\partial_x{\varphi}}\big]\big\} \Big)-V\\
&\Big(\big\{4 B^2 x^2+4 B^2 \alpha ^2 x^2+4 B^2 U (\partial_r{\psi})^2 x^2+4 B^2 \alpha ^2 U (\partial_r{\psi})^2 x^2\\
&+8 B^2 \alpha  U (\partial_r{\psi})^2 x^2-4 \sqrt{3} B U {\partial_y{\psi}} {\partial_r{\varphi}} {\partial_r{\psi}}\, x\\
&+3 U (\partial_y{\psi})^2 (\partial_r{\varphi})^2+3 (\partial_y{\varphi})^2 \big(U (\partial_r{\psi})^2+1\big)\\
&-4 \sqrt{3} B \alpha  U {\partial_y{\psi}} {\partial_r{\varphi}} {\partial_r{\psi}}\, x+4 B^2 y^2 \alpha ^2+4 B^2 y^2 \alpha ^2 U (\partial_r{\psi})^2\\
&+8 B^2 \alpha  x^2-4 \sqrt{3} B y \alpha  U {\partial_x{\psi}} {\partial_r{\varphi}} {\partial_r{\psi}}\\
&+{\partial_y{\varphi}} \big[4 B x (\alpha +1) \sqrt{3}+2 U {\partial_r{\psi}} \big(2 \sqrt{3} B x (\alpha +1) {\partial_r{\psi}}\\
&-3 {\partial_y{\psi}} {\partial_r{\varphi}}\big)\big]+3 U (\partial_x{\psi})^2 (\partial_r{\varphi})^2+3 (\partial_x{\varphi})^2 \big(U (\partial_r{\psi})^2+1\big)\\
&+{\partial_x{\varphi}} \big[4 B y \alpha  \sqrt{3} U (\partial_r{\psi})^2-6 U {\partial_x{\psi}} {\partial_r{\varphi}} {\partial_r{\psi}}+4 B y \alpha  \sqrt{3}\big]\big\}\psi ^2\\
&-\big\{4 \big[ y^2 \alpha ^2+x^2 (\alpha +1)^2\big] B^2+4 x (\alpha +1) \sqrt{3} {\partial_y{\varphi}} B+3 (\partial_y{\varphi})^2\\
&+4 y \alpha  \sqrt{3} {\partial_x{\varphi}} B+3 (\partial_x{\varphi})^2\big\} \psi ^4+3 \big\{ (\partial_y{\psi})^2+(\partial_x{\psi})^2\big\}\Big) \\
&+\frac{V}{U} \Big( \psi^{2} \big\{ 4B^{2}\big[y^{2}\alpha^{2} + x^{2} (1+ \alpha)^{2}\big] (\partial_{t} \psi)^{2} + 4\sqrt{3}B \partial_{t}  \psi  \\
& \big[y \alpha (\partial_{x}\varphi \partial_{t}\psi - \partial_{t}\varphi \partial_{x}\psi) + x (1+ \alpha)(\partial_{y}\varphi \partial_{t}\psi - \partial_{t}\varphi \partial_{y}\psi)\big]  \\
& + 3 \big[(\partial_{x}\varphi)^{2} (\partial_{t}\psi)^{2} - 2 \partial_{t}\varphi \partial_{x}\varphi \partial_{t}\psi \partial_{x}\psi \\
& + (\partial_{t}\varphi)^{2} (\partial_{x}\psi)^{2} + (\partial_{y}\varphi \partial_{t}\varphi \partial_{y}\psi)^{2}\big] \big\}  + 3 \big\{(\partial_{t}\varphi)^{2} (\psi^{2} - \psi^{4}) \\
&  + (\partial_{t}\psi)^{2} + \psi^{2} (\partial_{t}\varphi \partial_{r}\psi - \partial_{r}\varphi \partial_{t}\psi)^{2} U\big\} V \Big)\Big].
\end{split}\label{determinant}
\ee

\section{Shooting method to find the quasinormal frequencies}\label{shottingmeth}

Concretely, we numerically integrated \eqref{eompsir} from the horizon using \eqref{seriespsi} and \eqref{coefseriespsi} to provide boundary conditions near $r_h$ for an arrangement of complex values for $\omega$ and read the coefficient $\psi^{(-1)}$ from each of the resulting solutions. We then created interpolating functions Re$[\psi^{(-1)}]($Re$[\omega],$Im$[\omega])$ and  Im$[\psi^{(-1)}]($Re$[\omega],$Im$[\omega])$ to locate the quasinormal frequencies as the intersections of the contours traced by Re$[\psi^{(-1)}]=0$ and Im$[\psi^{(-1)}]=0$ in the $\lbrace$Re$[\omega],$Im$[\omega]\rbrace$ plane.

\end{document}